# Bicycle flow dynamics on wide roads: Experiment and modeling

Ning Guo[a], Rui Jiang[b,1], SC Wong[c,1], Qing-Yi Hao[d], Shu-Qi Xue[b], Mao-Bin Hu[e]

[a] *School of Automotive and Transportation Engineering, Hefei University of Technology, Hefei 230009, P.R. China*
[b] *Key Laboratory of Transport Industry of Big Data Application Technologies for Comprehensive Transport, Ministry of Transport,, Beijing Jiaotong University, Beijing 100044, P.R. China*
[c] *Department of Civil Engineering, The University of Hong Kong, Pokfulam Road, Hong Kong, P.R. China*
[d] *School of Mathematics and Computational Science, Anqing Normal University, Anqing 246133, P.R. China*
[e] *School of Engineering Science, University of Science and Technology of China, Hefei 230026, P.R. China*

**Abstract**

Cycling is a green transportation mode, and is promoted by many governments to mitigate traffic congestion. However, studies concerning the traffic dynamics of bicycle flow are very limited. This study experimentally investigated bicycle flow dynamics on a wide road, modeled using a 3-m-wide track. The results showed that the bicycle flow rate remained nearly constant across a wide range of densities, in marked contrast to single-file bicycle flow, which exhibits a unimodal fundamental diagram. By studying the weight density of the radial locations of cyclists, we argue that this behavior arises from the formation of more lanes with the increase of global density. The extra lanes prevent the longitudinal density from increasing as quickly as in single-file bicycle flow. When the density is larger than 0.5 bicycles/m$^2$, the flow rate begins to decrease, and stop-and-go traffic emerges. A cognitive-science-based model to reproduce bicycle dynamics is proposed, in which cyclists apply simple cognitive procedures to adapt their target directions and desired riding speeds. To incorporate differences in acceleration, deceleration, and turning, different relaxation times are used. The model can reproduce the experimental results acceptably well and may also provide guidance on infrastructure design.

## 1. Introduction

With economic development and population growth, especially in developing countries like China, urban transportation systems have come under huge pressure. The reliance on private cars for travel has had severe negative effects, such as traffic congestion, air pollution, and noise pollution. Government transportation departments now face the challenge of promoting environmentally-friendly and physically-active transportation modes. To this end, transportation management strategies often encourage walking and cycling for both commuting and short utilitarian trips (Mailbach et al., 2009; Rojas-Rueda et al., 2011). Walking and cycling have the potential to reduce air pollution and mitigate climate change, while offering individuals a low-cost travel option that benefits their personal health and fitness (Wen and Rissel, 2008;WHO, 2009, 2011).

Better understanding of every aspect of traffic behavior is vital for the design of urban infrastructures and traffic management. In the past decades, researchers have placed a significant focus on traffic dynamics as a way to study the fundamental characteristics of traffic flow. Abundant research findings on the traffic flow of motor vehicles and pedestrians are available. Conventional traffic flow theories are usually concerned primarily with motor vehicles. Many empirical/experimental studies and simulation models have been implemented to investigate vehicular flow dynamics, characterizing a number of traffic flow properties, such as oscillations, instability, and capacity drop (Chandler et al., 1958; Gazis et al., 1961; Kerner and Rehborn, 1997; Kerner, 1998; Mauch and Cassidy, 2002; Gazis, 2002; Nagel et al., 2003; Laval, 2006; Schönhof and Helbing, 2007, 2009; Treiber et al., 2010; Jiang et al., 2015; Tian et al., 2015, 2016). Meanwhile, in recent years, the study of pedestrian traffic flow has also drawn wide attention. A number of self-organization phenomena, such as the "faster-is-slower" effect, lane formation, and turbulent movement, have been observed and investigated (Navin and Wheeler, 1969; Muramatsu et al., 1999; Helbing et al., 2000, 2005; Hughes, 2002; Chattaraj et al., 2009; Moussaïd et al., 2011; Guo and Huang, 2012; Zhang et al., 2012). However, previous studies of cycling dynamics are scarce.

Cycling has gradually developed into an important transportation mode in urban life. Many governments have advocated cycling by adopting pro-bicycle measures such as bicycle-exclusive road networks and full integration with public transportation. Cycling can be faster than other forms of transport for short or even medium-distance trips (Zahabi et al., 2016). As a result, bicycles have become particularly important in some countries. For example, the percentage of bicycle trips among overall transport is 10% in Germany and Sweden, 11% in Finland, 18% in Denmark, and 27% in the Netherlands (Pucher and Buehler, 2008). In China, bicycle sharing is flourishing. According to data from the Scientific Research Institute of the Ministry of Transportation of China, congestion has decreased by 7.4%, 6.8% and 4.1% in Beijing, Shenzhen and Guangzhou, respectively, due to bicycle sharing[2].

To better design, operate, and control bicycle facilities, it is essential to understand the traffic dynamics of bicycle flow. However, most current studies of bicycle flow focus on issues such as safety (Dozza et al., 2016; Hamann and Peek-Asa, 2017), level of service (Allen-Munley et al., 2004; Wang et al., 2008; Elias, 2011; Dozza and Werneke, 2014), capacity (Highway Capacity Manual, 2000; Homburger, 1976; Botma and Papendrecht, 1991; Raksuntorn and Khan, 2003), flow rate versus density

---

[1] Corresponding authors.
  *E-mail address:* jiangrui@bjtu.edu.cn, hhecwsc@hku.hk

[2] https://www.chinadialogue.net/article/show/single/en/9853-Bike-sharing-schemes-Flourishing-or-running-riot- (accessed on 2018 Oct. 1)

(Zhang et al., 2013; Li et al., 2015; Jin et al., 2017; Navin, 1994), or the effect of bicycles on vehicle traffic flow (Zhao et al., 2013; Luo et al., 2015; Zhang et al., 2017). Studies concerning the traffic dynamics of bicycle flow are very limited.

Motivated by this fact, in our previous work (Mai et al., 2012; Jiang et al., 2017), we performed a single-file bicycle flow experiment on a 146 m circular road. The fundamental bicycle flow diagram was presented and the trajectories of each bicycle were extracted. We found that single-file flow exhibited a unimodal fundamental diagram, and traffic jams spontaneously formed above a critical density of $\rho \approx 0.37$ bicycles/m. Zhang et al. (2014) performed a similar single-file bicycle experiment on an 86 m circular track. They found that at $\rho = 0.3$ bicycles/m the flow rate dropped sharply, marking the transition to a congested state. Stop-and-go waves were observed when $\rho = 0.384$ bicycles/m. The critical value was very similar to Jiang et al. (2017). Zhao and Zhang (2017) also conducted single-file experiments for bicycles on a 52 m circular track, and compared single-file experiments involving cars, pedestrians, and bicycles. Their results were also very similar to those of Jiang et al. (2017) and Zhang et al. (2014). Moreover, many models have been proposed to reproduce and analyze the experimental findings (Navin, 1994; Andresen et al., 2013; Gould and Karner, 2009; Jia et al., 2007; Xue et al., 2007; Liang et al., 2012).

In the above mentioned single-file experimental studies, bicycles were not allowed to overtake each other. The allowed behavior was restricted to bicycle-following, which is not usually the case in real bicycle flow. On a real road, cyclists can follow, ride side-by-side, zipper or overtake the leading cyclist. The latter three behaviors are absent in a single-file experiment. To address more realistic cycling dynamics, we performed an experiment on a 3-m-wide track. In the experiment, participants were allowed to ride in their usual manner, and were not limited to bicycle-following. The experimental aim is to observe the flow rate at different global densities and to find the capacity of the bicycle flow. We also investigate the traffic dynamics of bicycle flows, including the potential emergence of stop-and-go traffic at different densities. It was found that the bicycle flow rate remained nearly constant across a wide range of densities. When the density was larger than 0.5 bicycles/m$^2$, the flow rate began to decrease, and stop-and-go traffic emerged.

In Jiang et al. (2017), a cellular automaton (CA) model was proposed to reproduce bicycle dynamics in single-file experiments. To model the bicycle flow on a wide road, one usually needs to generalize the CA model to the multilane situation by considering lane changing behavior. However, bicycles do not move strictly within lanes like other vehicles. Moreover, as our experimental track is circular, a CA model is inconvenient. In CA models of pedestrians, pedestrians are usually modeled as circles/square. A bicycle is not a circle, and it is difficult to describe the turning behavior of bicycles in a CA model. Therefore, the continuous model is a better choice for bicycle simulations. To account for the non-lane-based behavior of bicycle flow, here we propose a cognitive-science-based model, in which cyclists apply simple cognitive procedures to adapt their target directions and desired riding speeds. The simulation results are in good agreement with the experimental results.

The paper is organized as follows. The next section presents our experimental setup and experimental results. Section 3 proposes a model of bicycle flow. Section 4 presents the simulation results. Conclusions are given in Section 5.

## 2. Experiment

2.1 Experimental setup

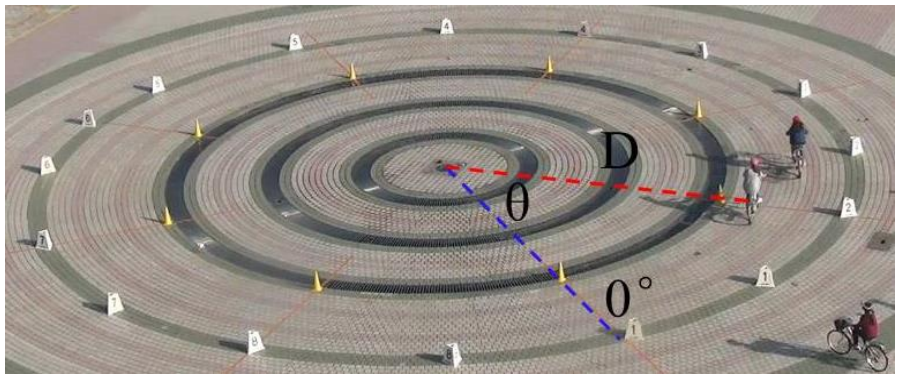

Fig. 1. Experimental field

The experiment was performed on May 7, 2016 at Anqing Normal University in China. The experiments were conducted on an artificial ring-shaped track, which are often used to study vehicle, bicycle, and pedestrian flow (Moussaïd et al., 2012; Tadaki et al., 2013). A periodic boundary is used to maintain the density of the system. This kind of setup removes disturbances from the density fluctuation, making it easier to find the critical values of the phase transition. The ring-shaped track is a common setup for implementing a periodic boundary. The inner radius and outer radius of the track were 8 m and 11 m respectively, and the boundaries were marked with dark floor tiles, warning signs, and number signs (Fig. 1). The area of the experimental track was thus approximately 180 m$^2$. The number signs partitioned the track into eight uniform subareas for statistical purposes. A total of 100 participants (undergraduates, 30 men, 70 women) took part in the experiments[3]. Each participant was assigned a serial number from 1 to 100, and they were requested to ride on the track in their usual manner. All of the participants were naïve to the purpose of the experiments. The maximum speed of a bicycle is around 15 km/h. In a post-experiment survey, all of the participants said that as the radius of the ring-shaped track was large, they felt that the difference between riding on the ring-shaped track and riding on a straight track was trivial.

---

[3] The experimental results are biased toward the behavior of young educated adults.

The full experimental procedure consisted of two rounds. Each round comprised 11 runs of experiments with different numbers of participants. Chronologically, the 11 runs involved the following numbers of participants: N = 50, 50, 40, 60, 30, 70, 20, 80, 10, 90, 100. The serial numbers were complementary between runs, as shown in Table 1. In each run of the experiments, the participants were initially distributed randomly on the track. They were requested to ride in the counter-clockwise direction. Each run lasts at least 3 minutes. As the density increases, the time increases to 5 minutes. In fact, we find that the bicycle flow reaches a stationary state. After the first round of experiments, all of the participants were allowed sufficient time to rest. Then, the second round was carried out. In the experimental process, we find that stop-and-go traffic emerges at the same density. Therefore, we did not carry out more runs.

Table 1. Serial numbers in each experimental run

| Run | Number of participants | Serial numbers |
| --- | --- | --- |
| 1 | 50 | 1~50 |
| 2 | 50 | 51~100 |
| 3 | 40 | 1~40 |
| 4 | 60 | 41~100 |
| 5 | 30 | 1~30 |
| 6 | 70 | 31~100 |
| 7 | 20 | 1~20 |
| 8 | 80 | 21~100 |
| 9 | 10 | 1~10 |
| 10 | 90 | 11~100 |
| 11 | 100 | 1~100 |

A video camera (SONY HDR-CX510E) was used to film the experiment from the roof of a classroom building neighboring the track. The horizontal distance from the camera to the center of the ring-shaped track is 53 m, and the vertical distance is 24 m. Cyclists' head were extracted by using Tracker software package. Although manual extraction poses some risk of errors, these should not have significantly impacted the experimental results.

2.2 Experimental results

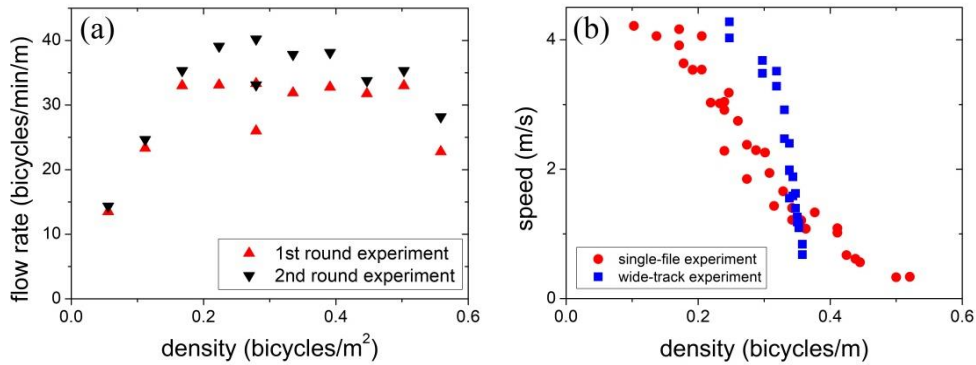

Fig. 2. Fundamental diagram. (a) Flow rate versus global density; (a) Speed versus density per lane.

We study the fundamental diagram of the flow rate versus the global density to determine the effect of the number of bicycles on system performance. The flow rate ($q$) is defined as the number of cyclists ($N_c$) crossing a line segment perpendicular to their direction of movement per unit time and per unit width.

$$q = N_c / \left[ (T - T_d) \cdot w \right]$$

(1)

Here $w = 3$ m is the width of the track. $T$ is the duration of the experiment. To discard the effect of transient time, the first $T_d = 0.5$ minutes of each experiment is excluded. We check different cross sections. The flow rate at different cross section is nearly the same.

Fig. 2(a) shows the fundamental diagram. One can see that when the density is smaller than 0.17 bicycles/m² (30 cyclists), the bicycles exhibit a free flow state, in which the flow rate increases with density. The flow rate then becomes almost constant until the density reaches 0.50 bicycles/m² (90 cyclists). In the 100-participant experiment, as the bicycle flow becomes congested, the flow rate begins to decrease. The flow rates in the second round of experiments are slightly larger than those in the first round. Table 2 shows the relative error in the flow rate between the two rounds as (second round-first round)/average of the two rounds. The maximum error is about 20%. One possible reason for this is that the participants were more familiar with the riding environment by the second round. The existence of a constant flow across a wide range of densities gives rise to a trapezoidal fundamental diagram, which is markedly different from the unimodal diagram of single-file bicycle flow (Jiang et al., 2017).

Table 2. Relative error in the flow rate between two rounds of experiment

| N | 10 | 20 | 30 | 40 | 50 | 60 | 70 | 80 | 90 | 100 |
|---|---|---|---|---|---|---|---|---|---|---|
| Relative error | 5.99% | 5.56% | 6.84% | 16.6% | 21.09% | 17.08% | 15.10% | 6.10% | 6.83% | 21.21% |

To study why the flow rate remains essentially constant across a wide range of densities, we plot snapshots of the location of (the head of) each cyclist in polar coordinates (Fig. 3). One can see that the bicycles are staggered so that the space between them can be utilized more efficiently on the wide track. Evidently, with the increase of the global density, the cyclists gradually form an increasing number of lanes. Specifically, one can see that all of the cyclists move around a polar radius of 9 m in the 10-bicycle experiment (Fig. 3(a)). When the number of bicycles increases, the cyclists segregate into two lanes around 9 m and 10 m polar radii, respectively (Fig. 3(b)). As the number of bicycles further increases, more lanes are formed (Fig. 3(c)). Of course, limited by the 3 m width of the track, the maximum number of lanes will be restricted.

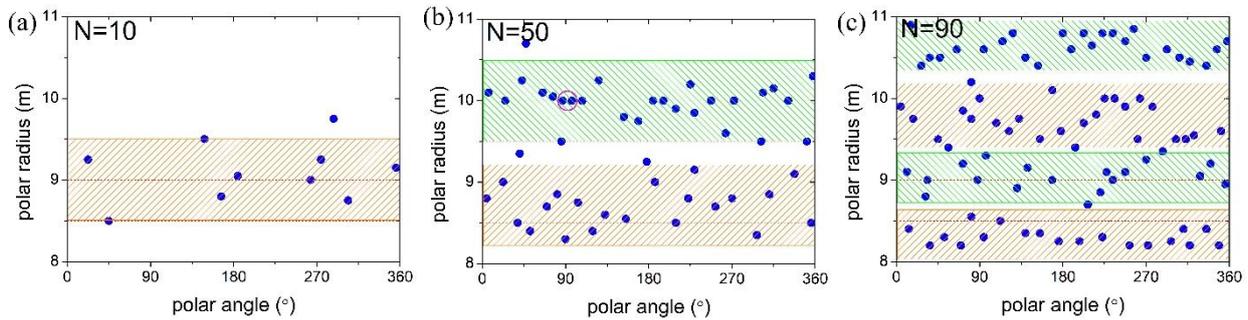

Fig. 3. Snapshots of cyclists' locations in polar coordinates. Note that some points seem very close to each other, such as the two points circled in panel (b). Actually, however, the difference in polar angle between the two points is 9.75°, and their distance is 1.7 m, exceeding the size of a bicycle. The shadowed rectangles roughly indicate the lanes

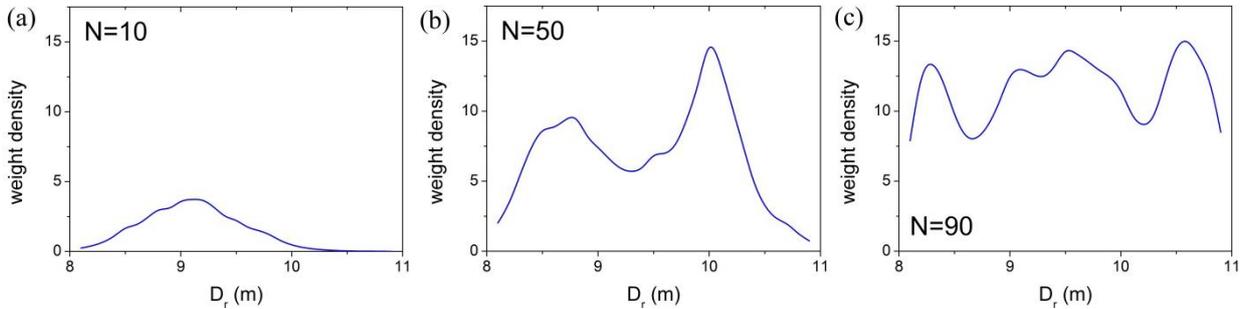

Fig. 4. Weight density in the polar radial direction

To quantify the above phenomenon, we introduce the weight density as proposed in Helbing et al. (2007)

$$f_w(D_r) = \sum_i e^{-\beta|D_i - D_r|} \qquad (2)$$

Here $D_i$ is the distance between cyclist $i$ and the track center, i.e. polar radial distance, see Fig. 1. $D_r$ is the distance between the two cyclists under consideration. $\beta$ is a parameter, and is set to 0.2. The index of summation is the cluster degree in the radial direction. Fig. 4 shows that with increasing global density, the number of peaks increases from 1 to about 5, which means that the number of lanes formed by bicycles increases. The linear fitting is *lane number*=0.227+0.0446*N*, $R^2$=0.956. Here lane number is defined as the number of peaks in the weight density function in the polar radial direction.

Using the lane number, we can calculate the density per lane and speed per lane. Compared to the experimental results for the single lane (Jiang et al., 2017), one can see that the speed per lane is larger in the wide-track experiment (Fig. 2(b)), indicating that although speed decreases as more bicycles occupy the track, cyclists respond by forming more lanes, which allows them to move side-by-side. As a result, the speed per lane increases. This utilization of space restrains the decrease of the flow rate, allowing it to remain essentially constant across a wide range of densities. The use of space affects lane formation in the unidirectional experiment, which is like the zipper effect in pedestrian flow in corridors. The zipper shape is helpful, not only to increase the distance between cyclists and thus decrease potential collisions, but also because it leads to lane formation. When the number of bicycles exceeds 90, formation of new lanes becomes difficult. As a result, the flow rate begins to decrease.

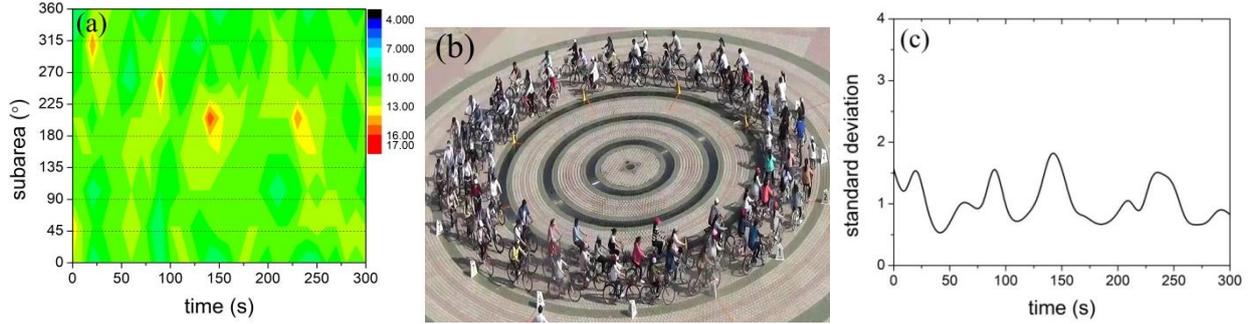

Fig. 5. Results of a 90-participant experimental run. (a) Evolution of the number of bicycles in the eight subareas, defined by their angle on the track; (b) Snapshot photograph; (c) Evolution of the standard deviation of the numbers of bicycles in the eight subareas. Dashed lines: boundaries of subareas.

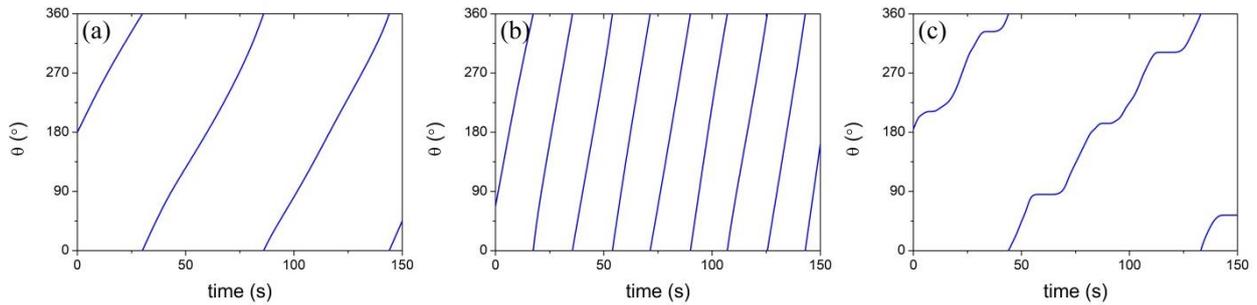

Fig. 6. Typical trajectory of the axial location of one cyclist in experiments involving (a) 90 participants, (b) 30 participants, (c) 100 participants.

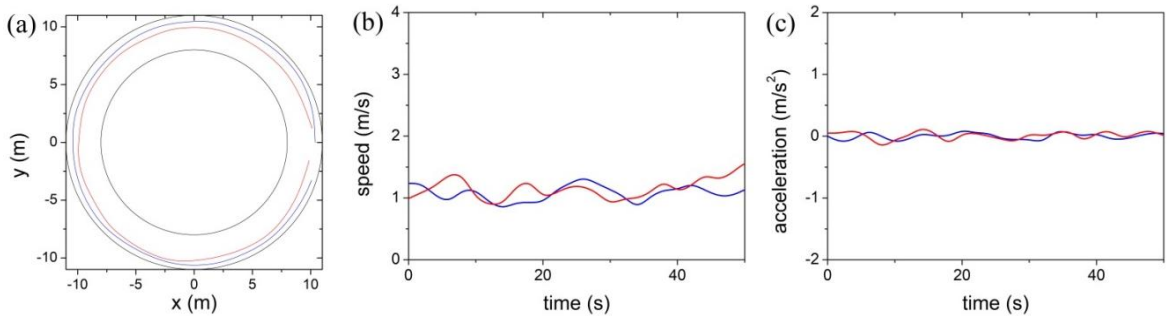

Fig. 7. Typical movement in 90-participant experimental run. (a) trajectories of two bicycles; (b) evolution of speeds; (c) evolution of accelerations.

Next, we study the evolution of bicycle flow at different densities. In the free flow state, the bicycles can move freely. This situation is very simple and does not require further illustration. We are mainly concerned with the saturation state and the congested state. To this end, we divide the track into eight equal areas. Now we present the experimental results of bicycle flow in the 90-participant experiment, corresponding to the highest density in the saturation flow state. After a short transient time, the cyclists are distributed fairly homogeneously (Fig. 5). Fig. 5 (a) shows the evolution of the number of cyclists in the eight subareas. Overall, the numbers of cyclists in each subarea are approximately the same at any one time (snapshot photograph in Fig. 5(b) and supplemental video S1). Randomness occasionally causes local jams (e.g., t = 90 s, t = 140 s in Fig. 5(a)), but the jams dissipate quickly. Fig. 5(c) describes the evolution of the standard deviation of the numbers of cyclists in the eight subareas over time. The standard deviation generally retains a small value around 1, which means that the bicycle flow is relatively homogeneous. Fig. 6(a) shows the typical trajectory of the axial location of a cyclist. The definition of axial location is shown in Fig. 1. One can see that the trajectory is roughly linear, which means that the bicycle has a constant riding speed.  Fig. 7 shows the typical movement. Because too many bicycles move on the track, few cyclists have space to overtake others (Fig.7(a)). The speed and acceleration keep constant basically, see Fig.7(b) and Fig.7(c).

Fig. 8(a) shows the evolution of the number of the cyclists in a 30-participant run, corresponding to the lowest density in the saturation state. Fig. 8(b) shows a typical snapshot photograph and Fig. 8(c) shows the standard deviation of the numbers of cyclists in the eight subareas. One can see that the distribution is again relatively homogeneous, and the small clusters that do form quickly dissipate. Fig. 6(b) shows that the typical trajectory is again linear. The overtaking can be observed at the low density (Fig.9). All other runs corresponding to the saturation state yielded similar results.

In contrast, in the congested flow (100-participant experiment), the cyclist distribution becomes nonhomogeneous, as shown in Fig. 10(a), in which a stop-and-go wave spontaneously emerges. As shown in Fig. 10(b), the local density in one area, indicated by the red box, is so large that the cyclists have to stop completely (see also supplemental video S2). The flow rate is thus a weighted average of "stop" state (~ 0) and "go" state (~ 0.55 bicycles/s/m). The traffic jam propagates steadily upstream along the centerline between the outer boundary and inner boundary, at a speed of approximately 1.5 m/s. The density in the jam is around $\rho_1$=0.8 bicycles/m$^2$. The density in the moving flow area is around $\rho_2$=0.45 bicycles/m$^2$, and the flow rate is $q_2$=0.55 bicycles/s/m. The propagation speed of the wave can be calculated by $q_2/(\rho_2-\rho_1)$, which yields a propagation speed of 1.57 m/s, close to the 1.5 m/s observed in the experiment. This mechanism used in vehicle traffic analysis can also be applied to describe the stop-and-go scenario in bicycle traffic. Fig. 10(c) shows that the standard deviation fluctuates around 3, much larger than that shown in Fig. 5(c) and 8(c). For the second run of the 100-participant experiment, similar results were observed. The propagation speed of traffic jams is again around 1.5 m/s. Fig. 6(c) and Fig.11(b) shows that due to the stop-and-go wave, the typical trajectory exhibits a zigzag profile, which is qualitatively different from that in the saturation state.

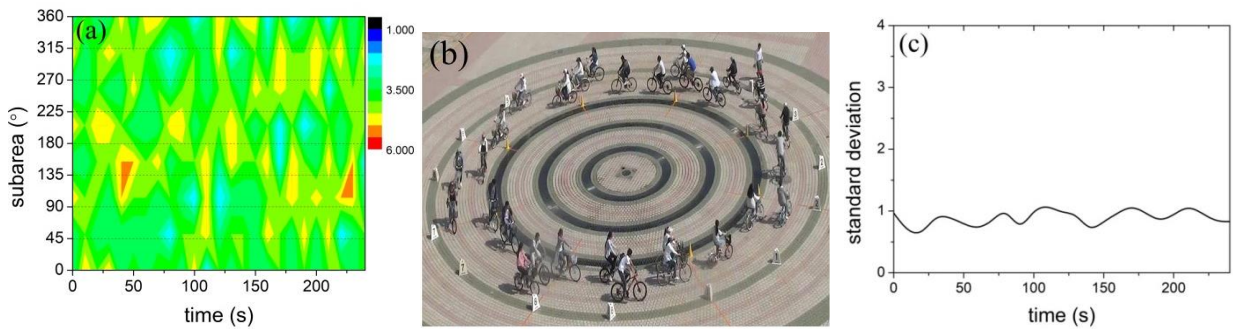

Fig. 8. Results of a 30-participant experimental run. (a) Evolution of the numbers of bicycles in the eight subareas; (b) Snapshot photograph; (c) Evolution of the standard deviation of the numbers of bicycles in the eight subareas. Dashed lines: boundaries of subareas.

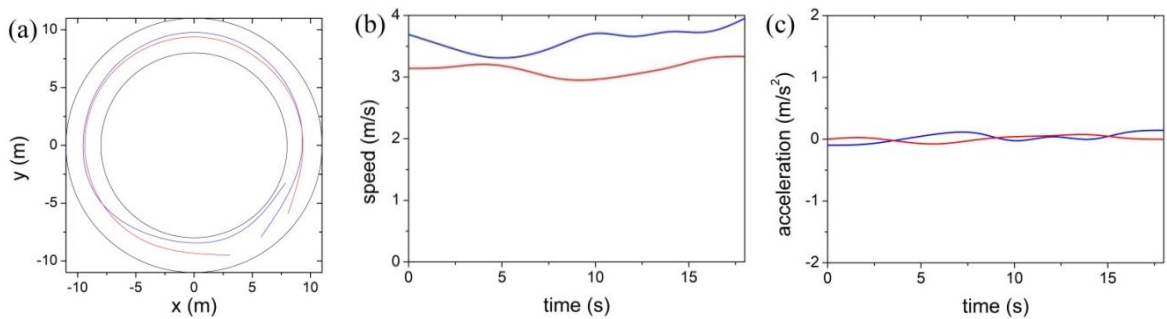

Fig. 9. Typical movement in 30-participant experimental run. (a) trajectories of two bicycles; (b) evolution of speeds; (c) evolution of accelerations.

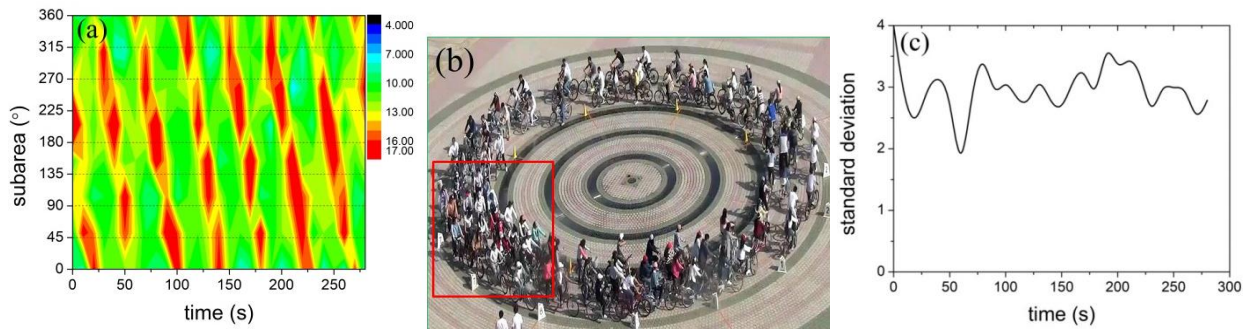

Fig. 10. Results of a 100-participant experimental run. (a) Evolution of the numbers of bicycles in the eight subareas; (b) Snapshot photograph; (c) Evolution of the standard deviation of the numbers of bicycles in the eight subareas. Dashed lines: boundaries of subareas.

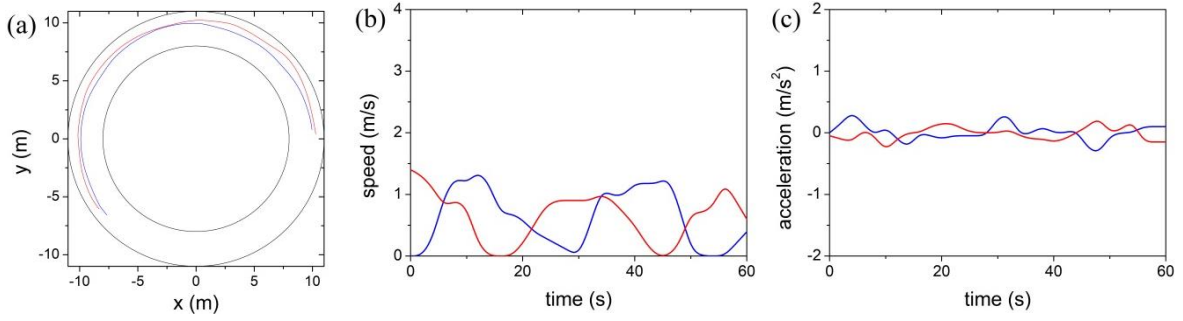

Fig. 11. Typical movement in 100-participant experimental run. (a) trajectories of two bicycles; (b) evolution of speeds; (c) evolution of accelerations.

Comparing bicycle, vehicle, and pedestrian flow, all three kinds of traffic exhibit stop-and-go patterns. However, at very high densities, pedestrian traffic can also undergo turbulent flow, which is absent in vehicle and bicycle traffic. This might be related to the fact that pedestrian flow at high densities is dominated by physical contact, which is absent in vehicle and bicycle traffic.

## 3. Model
3.1 Model formulation

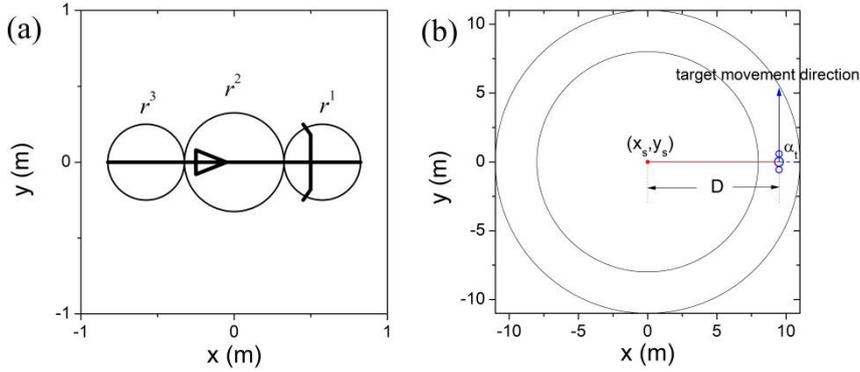

Fig. 12. (a) Simplified structure of bicycle/cyclist unit: three-circle shape. (b) target movement direction.

In this section we propose a cognitive-science-based model of bicycle flow, based on a model originally proposed for pedestrian flow (Moussaïd et al., 2011). We assume that the minimum comfortable space required by a cyclist is formed by three circles (Fig. 12(a); see also Thompson and Marchant (1995) and Qu et al. (2014), in which the pedestrians were modeled as three-circle-shaped). The bicycle does not have a regular structure and hence the bicycle is sometimes simplified to an ellipse. The three-circle shape is similar to an ellipse, but makes the distance to the collision easier to calculate. The radii are $r^1$ for the front circle, $r^2$ for the middle circle, and $r^3$ for the rear circle. When a cyclist's minimum comfortable space overlaps with that of another cyclist, the cyclist at the back will stop and wait for the cyclist in front to move forward.

We denote $\alpha_t$ as the angle of target movement direction, which is set to be the tangent direction of the ring (Fig.12(b))

$$\left(\cos(\alpha_t), \sin(\alpha_t)\right) = \left(-\frac{y - y_s}{D}, \frac{x - x_s}{D}\right)$$

(3)

where $x/x_s$ and $y/y_s$ are respectively the horizontal and vertical coordinates of the center of the bicycle/track. $D$ is the distance between the center of the middle bicycle and the track center.

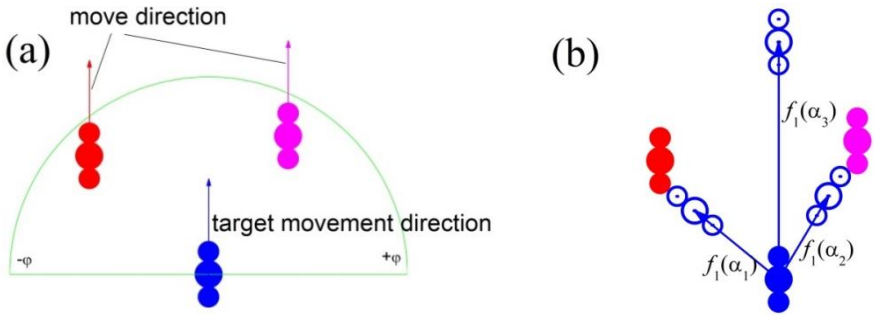

Fig. 13. (a) Initial location. (b) Three candidate directions. Hollow circles are predicted locations.

In the model, it is assumed that a cyclist chooses his or her movement direction in the range $[\alpha_t - \varphi, \alpha_t + \varphi]$. For all candidate directions $\alpha$ in this range, cyclist $i$ evaluates the distance to the first collision $f_1(\alpha)$, considering other cyclists' locations. For example, see Fig. 13. This model assumes that cyclist $i$ wishes to move at the target speed $v^{max}$ along the candidate direction $\alpha$, and cyclist $i$ assumes that the other cyclists do not move. As a result, given the candidate direction $\alpha$, cyclist $i$ will collide into another cyclist after a time interval $\Delta t$. For example, as shown in Fig. 13(b), cyclist $i$ will collide into another cyclist along candidate directions $\alpha_1$ and $\alpha_2$. In contrast, cyclist $i$ will not collide into any cyclist along direction $\alpha_3$. In the former two directions, $f_1(\alpha)$ is set to $f_1(\alpha) = v^{max} \cdot \Delta t$, denoting the distance to the first collision in direction $\alpha$. If no collision will occur in direction $\alpha$ (e.g., direction $\alpha_3$) or $f_1(\alpha)$ exceeds a default maximum value $d_{max}$, then $f_1(\alpha)$ is set to $f_1(\alpha) = d_{max}$.

For potential collisions with the track boundaries, $f_2(\alpha)$ is calculated by

$$f_2(\alpha) = \min\left[v^{max} \cdot \Delta t', d_{max}\right]$$

(4)

Here $\Delta t'$ is the time needed to collide with the boundary if the bicycle moves with speed $v^{max}$ along direction $\alpha$. Then, $f(\alpha)$ is defined as

$$f(\alpha) = \min\left[f_1(\alpha), f_2(\alpha)\right] \quad (5)$$

The chosen direction $\alpha_{des}$ is obtained through minimizing the function $d(\alpha)$

$$\alpha_{des}(t) = \arg\min[d(\alpha)]$$

(6)

in which $d(\alpha)$ is defined as
$$d^2(\alpha) = d_{max}^2 + f^2(\alpha) - 2d_{max}f(\alpha)\cos(\alpha_t - \alpha) \quad (7)$$

The desired speed is given by,

$$v_{des}(t) = \min\left[v^{max}, f(\alpha_{des})/\tau_1\right]$$

(8)

$\tau_1$ is the relaxation time. In our model, it is assumed that the bicycle direction (pointing from the center of the rear circle to that of the front circle) is identical to the direction of current velocity $\vec{v_i}$. Now we compare the two vectors $\vec{v_{des}}$ and $\vec{v_i}$. We split the vector $\vec{v_{des}}$ into two vectors, $\vec{v_{des1}}$ along the direction of $\vec{v_i}$, and $\vec{v_{des2}}$ vertical to $\vec{v_i}$. We assume that $\vec{v_{des1}}$ is due to acceleration (pedaling) or deceleration (e.g., cessation of pedaling, or braking), and $\vec{v_{des2}}$ is due to turning of the handlebar. Thus, when $|\vec{v_{des1}}| > |\vec{v_i}|$, the cyclist is accelerating, and when $|\vec{v_{des1}}| < |\vec{v_i}|$, the cyclist is decelerating. Therefore, the resulting acceleration equation is

$$\frac{d\vec{v_i}}{dt} = \begin{cases} \min\left(\frac{|\vec{v_{des1}} - \vec{v_i}|}{\tau_2}, a_a\right) \cdot \frac{\vec{v_{des1}}}{|\vec{v_{des1}}|} + \frac{\vec{v_{des2}}}{\tau_4}, & |\vec{v_{des1}}| \geq |\vec{v_i}| \quad (9a) \\ -\min\left(\frac{|\vec{v_{des1}} - \vec{v_i}|}{\tau_3}, a_d\right) \cdot \frac{\vec{v_{des1}}}{|\vec{v_{des1}}|} + \frac{\vec{v_{des2}}}{\tau_4}, & |\vec{v_{des1}}| < |\vec{v_i}| \quad (9b) \end{cases}$$

$\tau_2$, $\tau_3$, $\tau_4$ are relaxation times corresponding to acceleration, deceleration, and turning respectively. $a_a$ and $a_d$ are the maximum acceleration and maximum deceleration of the bicycle.

3.2 Sensitivity analysis

In the simulation, the size of the three-circle shape matches the size of real bicycle. The radii of the three circles are set to $r_i^1 = 0.25$ m, $r_i^2 = 0.325$ m, $r_i^3 = 0.25$ m. Based on physiological features related to safety engineering (He and Lin, 2000), the viewing range $\varphi = 90°$, and $d_{max} = 5$ m. According to mechanical characteristics of bicycles, the maximum acceleration is set to $a_a = 3$ m/s$^2$ and maximum deceleration is set to $a_d = 6$ m/s$^2$ [4]. Other parameters are shown in Table 3. We perform a sensitivity analysis of the four relaxation times. With the increase of $\tau_1$ or $\tau_2$, the cyclists require more time to accelerate to their desired speeds, resulting in a lower flow rate (Fig. 14(a) and (b)). When $\tau_3$ decreases, the cyclists can decelerate more quickly, also leading to a lower flow rate (Fig. 14(c)). $\tau_4$ influences the turning speed, which has a trivial effect on the flow rate (Fig. 14(d)).

---

[4] The coefficient of sliding friction between a bicycle wheel and the ground is 0.6-0.85 drawn from the table of friction coefficients: https://en.wikipedia.org/w/index.php?title=Friction&oldid=841213718 (accessed on 2018 Oct.1), so it is reasonable to set maximum deceleration to 6 m/s$^2$.

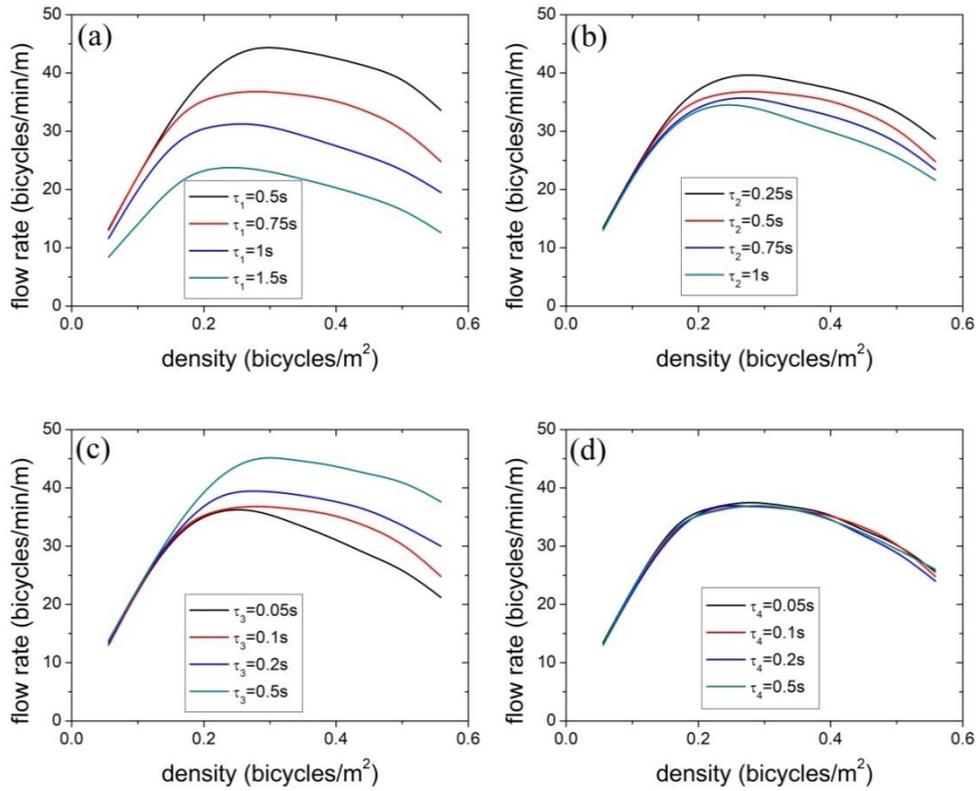

Fig. 14. The effect of (a) τ1, (b) τ2, (c) τ3, (d) τ4 on the flow rate.

3.3 Parameter calibration

The calibration method is drawn from Davidich and Köster (2012). The individual fitness is defined as

$$F = 1 \Big/ \sum_{i}^{10} \left| flow_i^{exp} - flow_i^{sim} \right|$$

(10)

Here $flow_i^{exp}$ is the average flow rate in the experiment, $flow_i^{sim}$ is the simulated flow rate at the same density. $i = 1\text{-}10$ corresponds to the 10 different global densities. A genetic algorithm (GA) is used for parameter calibration. The probability of selecting an individual is

$$P_l = F_l \Big/ \sum_{n} F_l$$

(11)

where $F_l$ is the fitness value of individual $l$, and $n$ is the number of individuals. The crossover rate is 0.5, and the mutation rate is 0.01. The initial population is distributed uniformly among the intervals, as shown in Table 3. The optimization process repeats 10 times, and the best set of values is selected. The calibrated values are listed in Table 3.

Table 3. Calibrated parameter values

| Parameter | Interval (Minimum/Maximum) | Value |
| --- | --- | --- |
| $v^{max}$ (m/s) | 0/10 | 4 |
| $\tau_1$ (s) | 0/2 | 0.75 |
| $\tau_2$ (s) | 0/2 | 0.5 |
| $\tau_3$ (s) | 0/2 | 0.1 |
| $\tau_4$ (s) | 0/2 | 0.1 |

## 4. Simulation results

4.1 Validation of simulation results on 3-m-wide road

Fig. 15 presents the relationship between density and flow rate. The RMSE between experimental and simulation results is 1.953 bicycles/min/m. The flow rate increases with density until the density reaches 0.22 bicycles/m$^2$ (40 bicycles), and then becomes nearly saturated at a value of around 35 bicycles/min/m. When the density exceeds 0.5 bicycles/m$^2$ (90 bicycles), the flow rate decreases sharply, and stop-and-go traffic emerges. The number of lanes also increases with the number of bicycles (Fig. 16). Linear fitting is *Lane number*=0.409+0.0373*N*, $R^2$=0.9382. The simulation results are in agreement with the experiment.

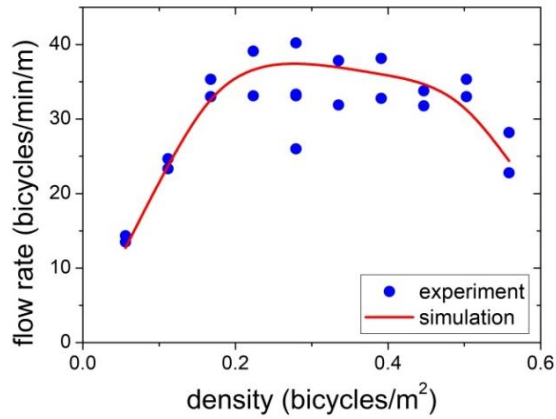

Fig. 15. Fundamental diagram obtained by simulation.

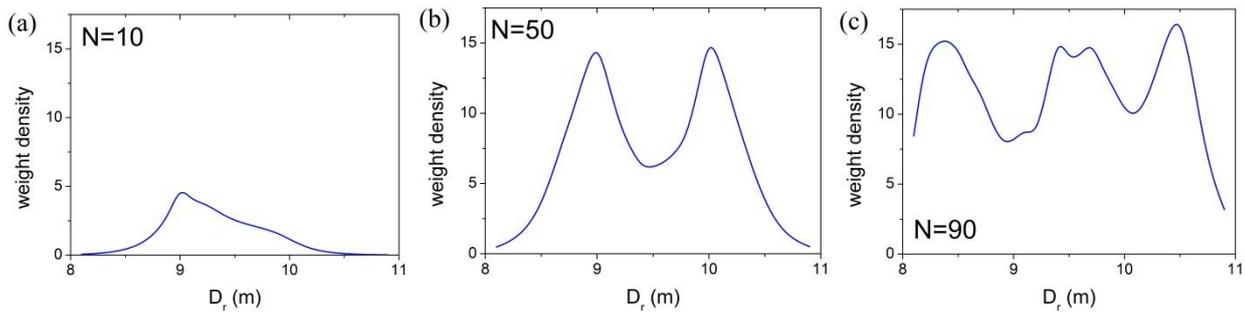

Fig. 16. Weight density in the radial direction obtained by simulation.

We compare the cyclist distribution between the simulated and experimental results. Fig. 17 shows a typical 90-bicycle simulation result. Local high densities and small jams occasionally emerge but dissipate very quickly (Fig. 17(a)). Fig. 17(b) presents a snapshot of the location of the bicycles, showing a relatively uniform distribution. The standard deviation of the numbers of bicycles in the eight subareas always remains small (Fig. 17 (c)). The trajectory is again roughly linear (Fig. 18(a)). The speed and acceleration are constant (Fig. 19).

The 30-bicycle simulation also closely reproduces the experimental results (Fig. 18(b), Fig. 20 and Fig. 21). Likewise, for all other densities in the saturation state, the simulation results are again consistent with the experiment.

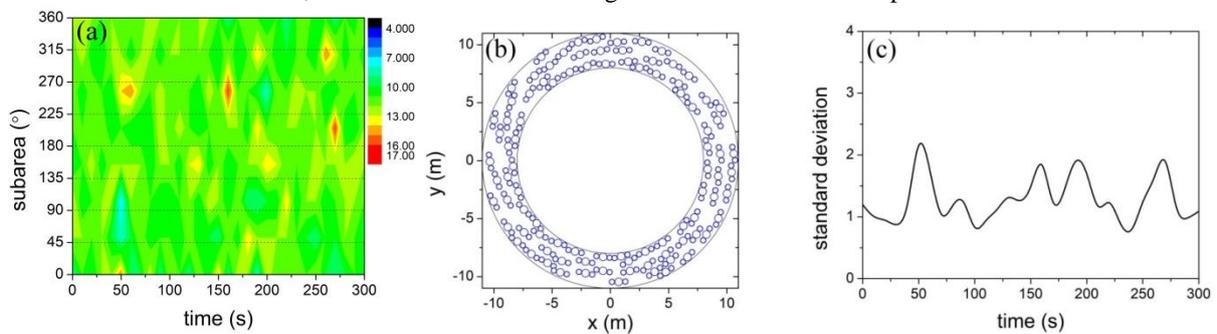

Fig. 17. Results of a 90-participant simulation run. (a) Evolution of the numbers of bicycles in the eight subareas; (b) Snapshot; (c) Evolution of the standard deviation of the numbers of bicycles in the eight subareas. Dashed lines: boundaries of subareas.

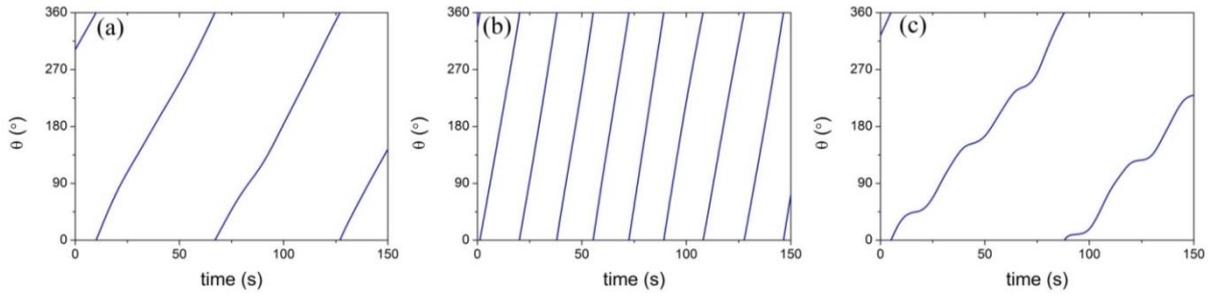

Fig. 18. Typical trajectory of the axial location of one cyclist in simulations involving (a) 90 participants, (b) 30 participants, (c) 100 participants.

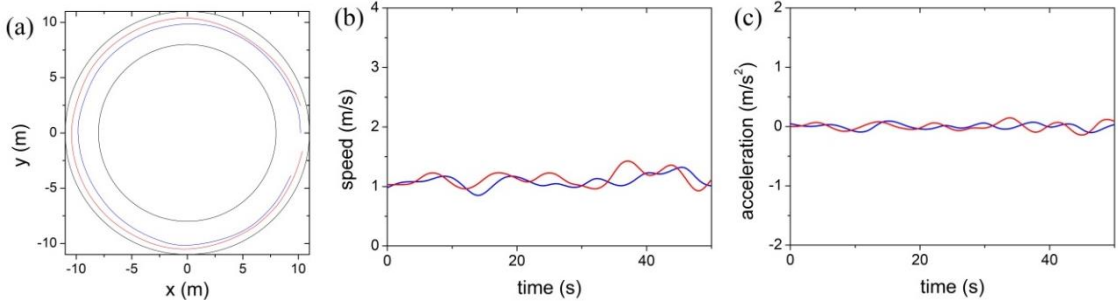

Fig. 19. Typical movement in a 90-participant simulation run. (a) trajectories of two bicycles; (b) evolution of speeds; (c) evolution of accelerations.

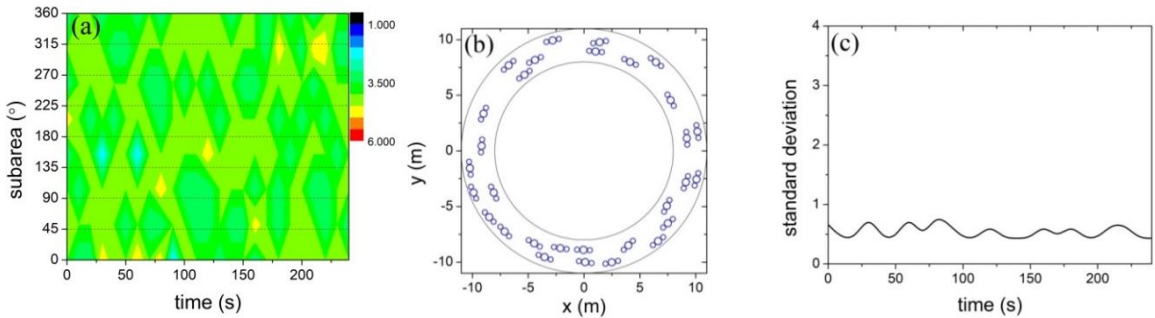

Fig. 20. Results of a 30-participant simulation run. (a) Evolution of the numbers of bicycles in the eight subareas; (b) Snapshot; (c) Evolution of the standard deviation of the numbers of bicycles in the eight subareas. Dashed lines: boundaries of subareas.

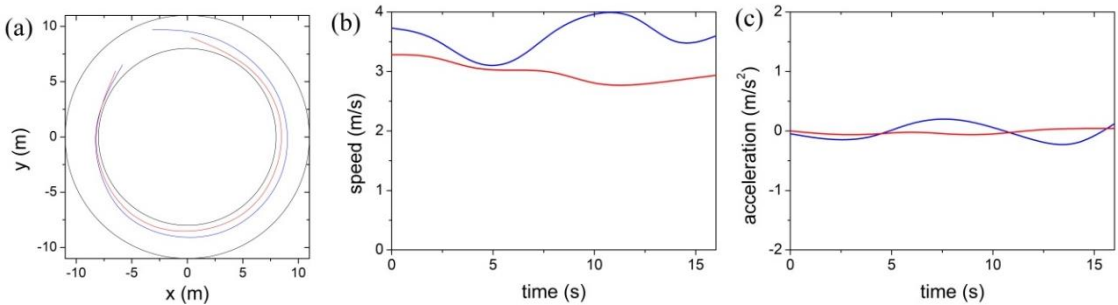

Fig. 21. Typical movement in a 30-participant simulation run. (a) trajectories of two bicycles; (b) evolution of speeds; (c) evolution of accelerations.

However, in the 100-bicycle simulation, a phase transition occurs (Fig. 22). From Fig. 22(a), the jam region spreads upstream, and stop-and-go traffic is observed (see also the zigzag trajectory in Fig. 18(c)). In Fig. 22(b), a nonuniform distribution emerges. The bicycles in the red box have stopped completely. Fig. 22(c) shows that the standard deviation of the numbers of bicycles in the eight subareas always remains large. The stop-and-go also occurs, see Fig.23.

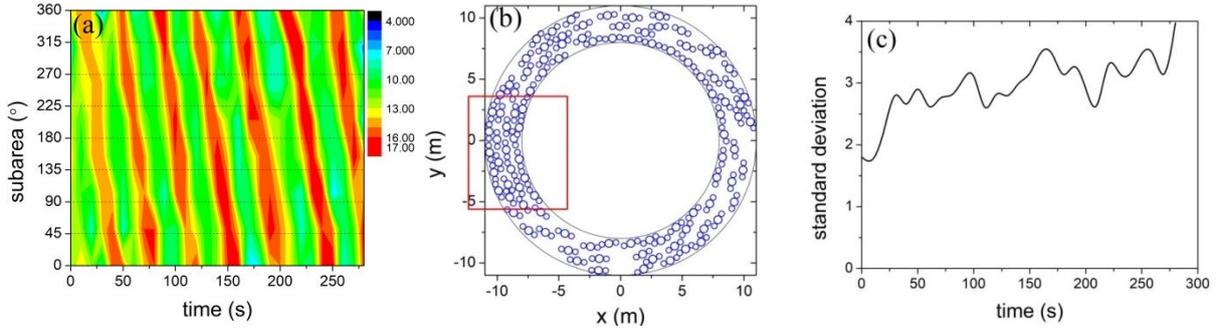

Fig. 22. Results of a 100-participant simulation run. (a) Evolution of the numbers of bicycles in the eight subareas; (b) Snapshot; (c) Evolution of the standard deviation of the numbers of bicycles in the eight subareas. Dashed lines: boundaries of subareas.

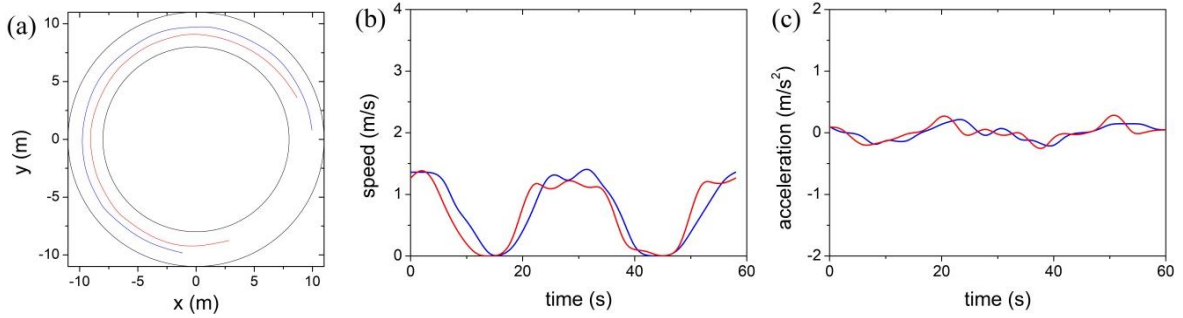

Fig. 23. Typical movement in a 100-participant simulation run. (a) trajectories of two bicycles; (b) evolution of speeds; (c) evolution of accelerations.

4.2 Validation of simulation in single-file bicycle flow

To simulate single-file bicycle flow, the model only needs to include longitudinal movement. Turning behavior need not be considered. This greatly simplifies the model, and the acceleration reduces to the following.

$$\frac{dv_i}{dt} = \begin{cases} \min\left(\left|\frac{v_{des} - v_i}{\tau_2}\right|, a_a\right), & v_{des} \geq v_i \quad (12a) \\ -\min\left(\left|\frac{v_{des} - v_i}{\tau_3}\right|, a_d\right), & v_{des} < v_i \quad (12b) \end{cases}$$

$$v_{des}(t) = \min\left[v^{\max}, \min(d, d_{\max})/\tau_1\right] \quad (13)$$

In the one-dimensional condition, the bicycle length is the sum of the three circles' diameters, set as 1.65 m. The view range $d_{\max}$= 5 m. $d$ is the distance between bicycles. The maximum acceleration $a_a$ = 3 m/s² and maximum deceleration $a_d$ = 6 m/s². The other parameters are obtained by calibration as in the previous subsection, and the values used in the simulations are listed in Table 4.

Table 4. Parameter values used in the simulations

| Parameter | Value |
| --- | --- |
| $v^{\max}$ (m/s) | 4 |
| $\tau_1$ (s) | 0.75 /1 |
| $\tau_2$ (s) | 0.5 |
| $\tau_3$ (s) | 0.1 |

Fig. 24 shows that the simulation results of the fundamental diagram are in agreement with the experiment. Fig. 25 shows the space-time diagrams at three global densities. When the density is smaller than the critical density of 0.37 bicycles/m, the bicycle flow is relatively homogeneous. Even when the simulation is started from a jam, the jam soon dissipates (Figs. 25(a) and (b)). When the density exceeds 0.37 bicycles/m, jams cannot dissipate (at least within the simulated time) (Fig. 25(c)). This is consistent with earlier experiments (Jiang et al., 2017).

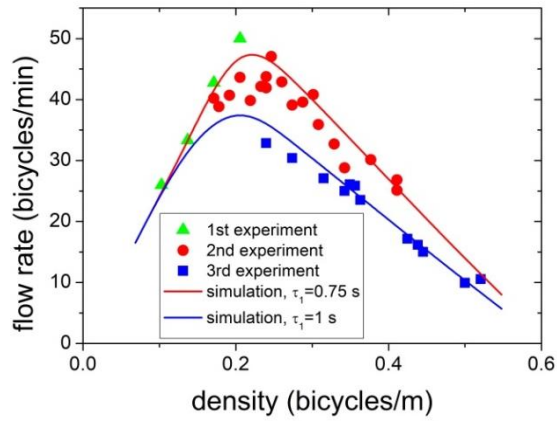

Fig. 24. Fundamental diagram of the single-file experiment reported in Jiang et al. (2017).

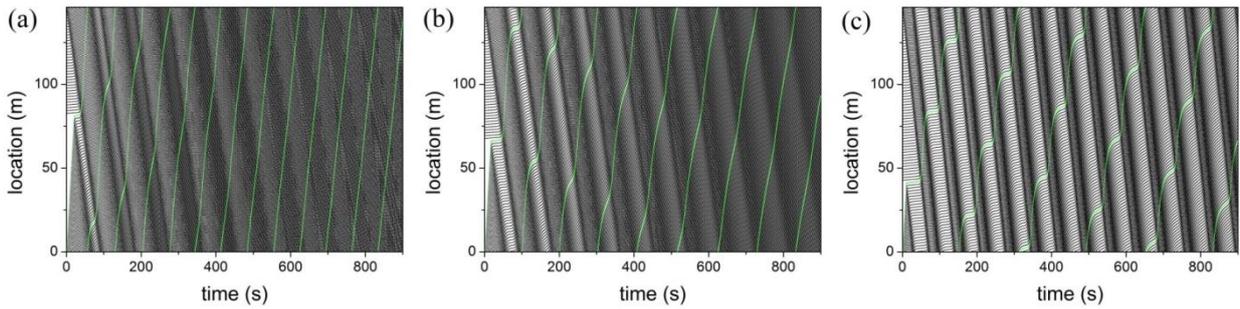

Fig. 25. Space-time diagram at global densities of (a) 0.267 bicycles/m (39 bicycles), (b) 0.329 bicycles/m (48 bicycles), (c) 0.432 bicycles/m (63 bicycles). $\tau_1 =$ 1 s in (a) and (c) and $\tau_1 = 0.75$ s in (b). To guide the eye, the green lines show the trajectory of one bicycle.

4.3 Discussion

Fig. 26(a) shows the probability density distribution of acceleration in the simulation. Here acceleration/deceleration are defined as the first term on the right-hand side of Eq. 9(a)/Eq. 9(b), respectively. The peak probability density emerges around 0 m/s$^2$, representing the situation in which the bicycles on average move with a constant speed. With the increase of global density, the peak value decreases and the distribution becomes wider, indicating that the bicycle flow gradually becomes less homogeneous. Fig. 26(b) shows the probability density distribution of the turning acceleration/deceleration in the simulation, which are defined as the second term on the right-hand side of Eq.9(a)/Eq.9(b). When the density is low, the cyclists tend to continually turn left because of the counter-clockwise direction of the circular track. With increasing density, there are more interactions among cyclists, and left turning is suppressed.

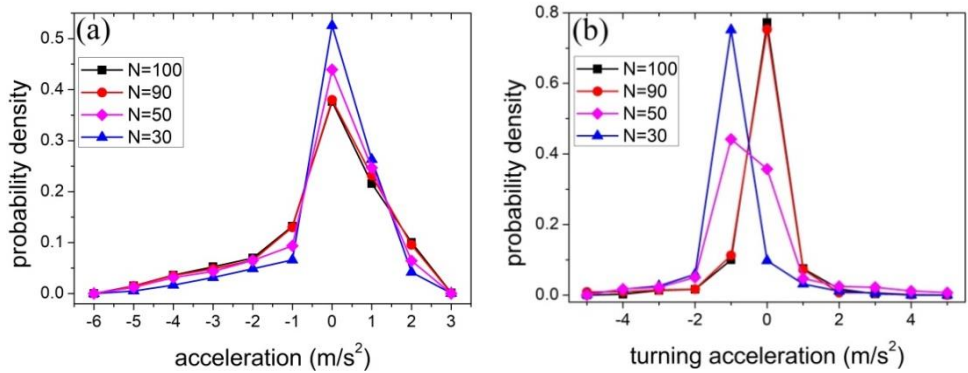

Fig. 26. Probability density distribution of (a) acceleration/deceleration (b) turning acceleration/deceleration, +/- indicates right/left turning.

Fig. 27(a) shows the effect of track width on the bicycle flow, in which the average value of both the inner radius and outer radius of the track is fixed at 9.5 m. It is found that when the track is widened from 2 m to 4 m, the flow rate is almost independent of the track width when the density is smaller than $\rho_{c1} \approx 0.15$ bicycles/m$^2$. When the density exceeds $\rho_{c1}$, the flow rate increases with the track width. The most likely explanation is that when the density is larger than $\rho_{c1}$, the boundary plays a nontrivial role. With increasing track width, the proportion of cyclists at risk of colliding into the boundary decreases. Therefore, the flow rate increases. When the track width exceeds 4 m, the boundary effect becomes negligible. Now we increase/decrease the inner radius and outer radius of the track, with the track width fixed at 3 m. As shown in Fig. 27(b), the radius of the boundaries has no significant impact on the flow rate.

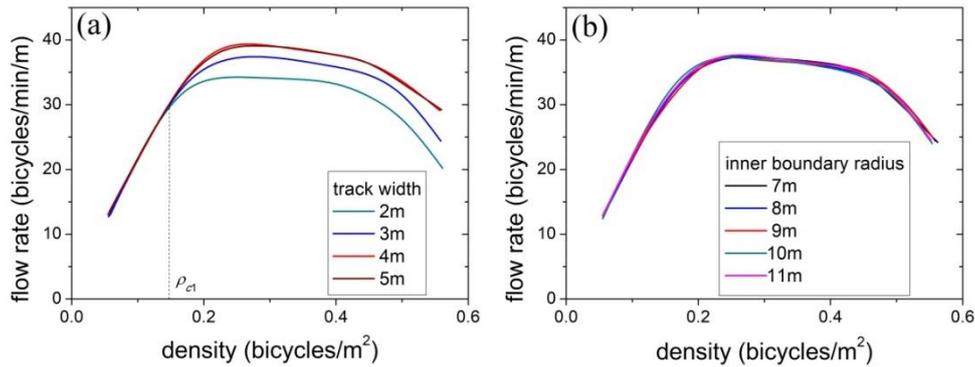

Fig. 27. Flow rate versus density at (a) different track width, with the average value of both the inner radius and outer radius of the track fixed at 9.5 m; (b) different inner boundary radius, with the track width fixed at 3 m.

The electric bicycle is an emerging alternative to the pedal bicycle. An important characteristic of electric bicycles is their higher maximum speed. Fig. 28(a) shows the relationship between flow rate and density at different maximum speeds of the bicycles. Higher maximum speeds lead to higher flow rates at low densities. However, when the density is above 0.5 bicycles/m$^2$, the flow rates remain almost constant even when the maximum speed exceeds 4 m/s. Finally, Fig. 28(b) presents the fundamental diagram for a range of bicycle sizes. At very low densities, bicycle size hardly affects the flow rate. However, larger bicycle sizes reduce the flow rate when the density is larger than 0.1 bicycles/m$^2$.

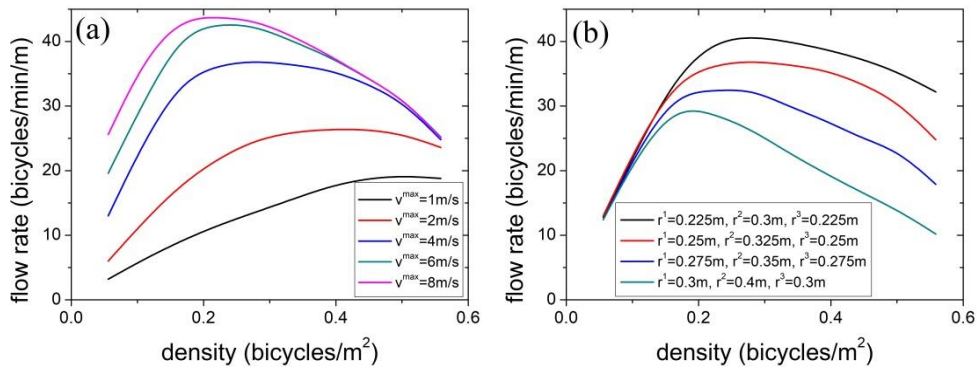

Fig. 28. Flow rate versus density (a) at different maximum speeds. (b) different bicycle sizes.

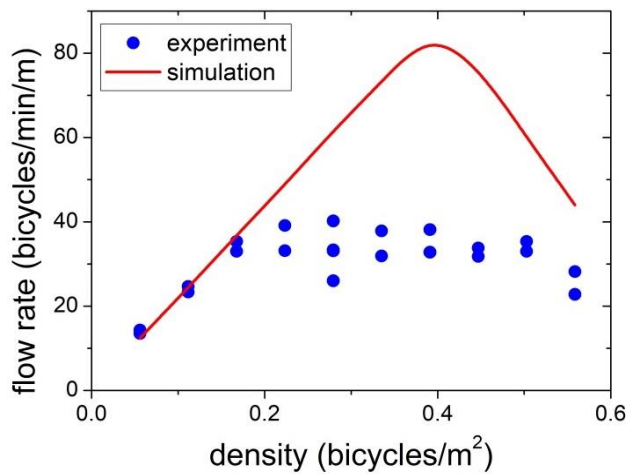

Fig. 29 Fundamental diagram. In the simulation, a cyclist assumes that the speed and direction of other cyclists' movements do not change.

Finally, we examine the assumption in our model concerning evaluation of $f_1(\alpha)$. In our model, to evaluate $f_1(\alpha)$, cyclists assume that other cyclists do not move. This is because bicycles only need a short stopping distance, particularly when it is congested. Therefore, it is rational to regard other bicycles as motionless. In contrast, if assuming that the speed and direction of other cyclists' movements do not change, a cyclist can follow the cyclist in front of her/him at a high speed even under a high-density condition, leading to too large flow rate in the medium/high density condition, see Fig. 29.

## 5. Conclusions

In this paper, we have experimentally studied the unidirectional bicycle flow on a 3-m-wide ring-shaped track. Experiments have been performed at 10 global densities. The fundamental diagram and the bicycle dynamics have been investigated. The flow rate increases with increasing density within the small-density range, and then remains nearly constant across a wide range of densities. The fundamental diagram thus exhibits a trapezoidal shape. We argue that this behavior arises from the formation of more lanes with the increase of global density. The extra lanes prevent the longitudinal density from increasing as quickly as in single-file bicycle flow. Moreover, a critical density of 0.5 bicycles/m$^2$ is found. Below the critical density, bicycles can move relatively uniformly. Above the critical density, a stop-and-go wave spontaneously emerges.

To account for the non-lane-based behavior of bicycle flow, we propose a cognitive-science-based model, in which cyclists apply simple cognitive procedures to adapt their target directions and desired riding speeds. Each bicycle is modeled as a three-circle shape for simplicity, and different relaxation times of acceleration, deceleration, and turning are considered. The simulation results are in acceptably good agreement with the experiment. Further, the effects of relaxation times, road width, inner and outer radius, maximum speed, and bicycle size have been investigated via simulation.

The quantitative effect of the fact that cyclists can look further ahead is a worthwhile topic for future studies. It is possible that the wave speed might change if cyclists reacted more slowly when they could not see what was ahead. We will test this in future studies by putting a wall along the track to restrict the participants' field of vision. Moreover, we will carry out bicycle experiment on a straight track, and compare the differences with a ring-shaped track. To further verify our model, we will experimentally study bicycle flow on roads with different widths.

**Acknowledgments**

This work is supported by the National Key R&D Program of China (No. 2017YFC0803300), the National Natural Science Foundation of China (Grants No. 71801066, 71621001, 71631002), and the Beijing Natural Science Foundation (Grant No. 9172013). The third author was also supported by the Francis S Y Bong Professorship in Engineering.